\documentclass[aps,pre,twocolumn,showpacs,superscriptaddress,groupedaddress]{revtex4-1}
\usepackage{graphicx}
\usepackage{dcolumn}
\usepackage{bm}

\usepackage{amssymb}
\usepackage{epsfig}
\usepackage{amsmath}
\usepackage{subfigure}
\usepackage{textcomp}
\usepackage{url}
\usepackage{float}
\usepackage{color}
\usepackage{cases}
\usepackage[normalem]{ulem}

\usepackage[colorlinks=true, urlcolor=blue, anchorcolor=blue, citecolor=blue,filecolor=blue,linkcolor=blue,menucolor=blue
]{hyperref}

\newcommand{\OO}{\mathcal{O}}

\newcommand{\pd}[2]{\frac{\partial #1}{\partial #2}}

\begin{document}

\preprint{APS/123-QED}

\title{Escape from a metastable state in non-Markovian population dynamics}

\author{Ohad Vilk $^{a,b,c}$}
\email{ohad.vilk@mail.huji.ac.il}
\author{Michael Assaf $^{a}$}
\email{michael.assaf@mail.huji.ac.il}
 \affiliation{$^a$ Racah Institute of Physics, The Hebrew University of Jerusalem, Jerusalem 91904, Israel,}
 \affiliation{$^b$ Movement Ecology Lab, Department of Ecology, Evolution and Behavior, Alexander Silberman Institute of Life Sciences, Faculty of Science, The Hebrew University of Jerusalem, Jerusalem 91904, Israel,}
 \affiliation{$^c$ Minerva Center for Movement Ecology, The Hebrew University of Jerusalem, Jerusalem 91904, Israel}


\begin{abstract}
     We study the long-time dynamics in non-Markovian single-population stochastic models, where one or more reactions are modelled as a stochastic process with a fat-tailed non-exponential distribution of waiting times, mimicking long-term memory. We focus on three prototypical examples: genetic switching, population establishment and population extinction, all with non-exponential production rates. The system is studied in two regimes. In the first, the distribution of waiting times has a finite mean. Here, the system approaches a (quasi)stationary steady state at long times, and we develop a general WKB approach for these non-Markovian systems. We derive explicit results for the mean population size and mean escape time from the metastable state of the stochastic dynamics. In this realm, we reveal that for sufficiently strong memory, a memory-induced (meta)stable state can emerge in the system.  
     In the second regime, the waiting time distribution is assumed to have an infinite mean. Here, for bistable systems we find two distinct scaling regimes, separated by an exponentially long time which may strongly depend on the initial conditions of the system.  
\end{abstract}

\maketitle

\section{Introduction} \label{sec:intro}
Stochastic populations, characterized by a mesoscopic number of interacting agents, often reside near an attractor, with occasional random fluctuations around it. These fluctuations, although infrequent, can have significant implications, such as triggering a transition to a different attractor, or to an absorbing state. This phenomenon, generally known as population escape, is of considerable interest across various fields, including physics, ecology, epidemiology, and biochemistry~\cite{bartlett1960stochastic, boyce1992population,  hanski2004ecology, chan2008paths,  assaf2011determining, vilk2020extinction,  hindes2022outbreak, boukanjime2020dynamics, nguyen2021environmental}.

In ecology, population escape from a long-lived metastable state plays an important role in the long-term stability of populations. Here, even a small population lingering on the brink of extinction can experience a rare fluctuation, securing its survival. On the other hand, a long-lived established population can undergo extinction. In both cases, escape occurs due to demographic noise, emanating from the stochastic nature of the reactions and discreteness of individuals~\cite{beissinger2000ecological, lande2003stochastic, kamenev2008colored, leisner2008stochastic, dykman2008disease,  levine2013impact,  vilk2018population,mendez2019demographic}. Population escape is also highly relevant for studying gene regulation and genetic switching, which describes the process of cells transitioning between distinct phenotypic states~\cite{ackers1982quantitative,elowitz2002stochastic, hornos2005self,raj2008nature, raser2005noise, munsky2012using, biancalani2015genetic,weinreich2005rapid, bressloff2017stochastic}. Notably, even in the absence of a driving signal, stochastic fluctuations of mRNA and proteins during the gene expression process can play an essential role in determining the stability of these states, and the corresponding transition rates~\cite{assaf2011determining}.

Stochastic population escape requires that the reaction rates be nonlinear in the population size and includes casting the problem into a first-passage problem of an underlying Markov process~\cite{gardiner1985handbook}. The mean time to escape can be found analytically in some cases, by using various approximations, see Refs.~\cite{kessler2007extinction, assaf2010extinction, assaf2017wkb, dobramysl2018stochastic} and references therein.  
The vast majority of studies dealing with population escape have assumed that the various reactions governing the escape process are Markovian, i.e., the inter-reaction times between any two events are exponentially distributed~\cite{yoshimura1996evolution, aurell2002epigenetics, elgart2004rare, escudero2004extinction, park2017extinction, shahrezaei2008analytical,  elgart2004rare, assaf2008noise,  gardiner1985handbook, tadokoro2020noise}.
Yet, it is known that many natural processes exhibit long delays or non-exponential intrinsic waiting times (WT) between reactions~\cite{ortega2004role, jia2011intrinsic, gurevich2013instabilities,  jo2014analytically, leier2015delay, schwartz2015noise, zhang2021analysis,  macias2020simple}. For example, single molecule tracking has shown that a nonergodic process governs the diffusion properties of plasma in the membrane of living cells~\cite{weigel2011ergodic}. Similar mechanisms were also found in different biological systems~\cite{jeon2011vivo, song2018neuronal} and in movement patterns of birds~\cite{vilk2022ergodicity, vilk2022classification}.    

Here we use the framework developed in Ref.~\cite{aquino2017chemical}, rooted in the continuous-time random walk (CTRW) theory~\cite{klafter1980derivation, metzler2000random, metzler2014anomalous}, to study population escape due to stochastic fluctuations in a generic single-population model with birth-death dynamics. We assume that the WTs for any birth event, whose underlying frequency scales nonlinearly with the population size, are distributed according to a fat-tailed distribution, leading to prolonged periods without birth events. This is a simplified mechanism that allows mimicking long-term memory, relevant in both ecological and biological processes~\cite{weigel2011ergodic, vilk2022ergodicity, fagan2013spatial}. While a similar model with linear rates was studied in Ref.~\cite{vilk2023non}, in the current manuscript we focus on nonlinear rates and study the effect of these prolonged periods on the mean time to escape from a metastable state.  We note that our model is similar to the model for genetic molecular switching studied in~\cite{yin2021optimal}, with several crucial differences. First, we generalize the model beyond gene regulation and consider a general birth-death model with an arbitrary birth rate. Second, while in~\cite{yin2021optimal} the WTs were drawn from an Erlang distribution, here we study intrinsic WTs drawn from a power-law distribution, which was shown to drive several ecological processes as discussed above. Third, we extend the formalism to study cases where the dynamics are nonstationary and nonergodic.  Finally, we apply a perturbation technique and analytically compute the escape times.

The remainder of the paper is organized as follows. In Sec.~\ref{SecProtein} we define the microscopic dynamics and obtain the chemical master equation. Next, in Sec.~\ref{sec:largealpha} we study the case of a WT distribution with a convergent mean, $\alpha >1$ (see below), while in Sec.~\ref{sec:smallalpha} we study the case of a divergent mean, $\alpha<1$, leading to nonstationarity at long times. In Sec.~\ref{conclusions} we summarize our results and discuss more complex models where long delays are expected to play an important role. Note that the numerical simulations we have implemented are done via a modified Monte-Carlo (MC) algorithm recently developed for non-Markovian stochastic systems~\cite{masuda2018gillespie}.

\section{microscopic dynamics and chemical master equation} \label{SecProtein}
Our starting point is a generic model of birth and death reactions, where these are respectively given by
\begin{equation} \label{reactions}
    n \rightarrow n+1 \; , \;\; n \rightarrow  n-1. 
\end{equation}
In general, the first reaction represents \textit{birth} or \textit{creation} of an agent, while the second reaction represents  \textit{death} or \textit{degradation} of an agent. For example, in  gene expression the reactions in Eq.~(\ref{reactions}) can represent translation and degradation (e.g., due to cell division) of proteins.

Such reactions are usually associated with an exponential rate in a point process. Yet, in a process with delays due to intrinsic WTs, one cannot assign a single rate for each reaction but rather a distribution of WTs~\cite{aquino2017chemical}. We thus define $\psi_1(\tau)$ as the WT distribution for a single birth event, $n \rightarrow n+1$, to occur between times $t$ and $t+\tau$, and $\psi_2(\tau)$ as the WT distribution for a single death event, $n \rightarrow n-1$, to occur between times $t$ and $t+\tau$.   
For concreteness, we  consider the following WTs
\begin{equation} \label{psi1}
     \psi_1(\tau) = \frac{\tilde{\kappa}(n) }{ [\tilde{\kappa}(n) \tau/\alpha +1]^{1+\alpha}}\; ;  \;\;\;
     \psi_2(\tau) = n e^{-n \tau }.  
\end{equation} 
The WT distribution for any death event to occur, $\psi_2(\tau)$, is exponential and can be directly derived by assuming that the probability distribution for the intrinsic WT follows a Poisson process with normalized rate 1~\cite{gardiner1985handbook}. 
In contrast, the WT distribution for a birth event, $\psi_1(\tau)$,
follows a power law, where the shape parameter $\alpha$ is the power law exponent and the scale of the distribution -- $\tilde{\kappa}(n)$ -- depends on the number of products $n$. Below, we discuss different choices of $\tilde{\kappa}(n)$ that correspond to systems where an escape from a locally metastable state is possible~\cite{assaf2011determining,assaf2010extinction,assaf2017wkb}. Importantly, for $\alpha<1$ the mean of the WT distribution $\psi_1$ diverges, while for $\alpha > 1$ the mean converges (in general, for $k< \alpha < k + 1$, only the $k$th moment and below are finite). As $\alpha$ controls the shape of the distribution of prolonged WTs we treat it here as a form of memory, where higher values of $\alpha$ entail weaker memory. In the limiting case of
$\alpha \to \infty$ the leading term of the birth rate is exponential and the process displays no long-term memory. 

Using the CTRW formalism developed in Ref.~\cite{aquino2017chemical} we write the following generalized chemical master equation for the probability distribution function (PDF) $P_n(t)$ of having $n$ agents at time $t$, see Appendix for details: 
\begin{equation} \label{MasterEquation}
    \frac{dP_n}{dt} = (E_n^1 -1) n P_n(t)  + (E_n^{-1} -1)\! \int_0^t\!\! M(n, t-t') P_{n}(t') dt', 
\end{equation}
where the step operators are defined by $E_k^j f(k) = f(k + j)$.
The kernel $M(n, t)$ is defined in terms of its Laplace transform, given by (see Appendix):
\begin{equation} \label{memorykernelexplicit}
    \tilde{M}(n, s) = \tilde{\kappa}(n) \frac{  E_{\alpha +1}\left(\alpha\frac{n+s }{\tilde{\kappa}(n) }\right)}{E_{\alpha}\left(\alpha\frac{n + s}{\tilde{\kappa}(n)
   }\right)}. 
\end{equation}
Here, $\tilde{M}(n,s)$ is defined to be the Laplace transform of $M(n, t)$, $s$ is the Laplace variable, and $E_m(z) \equiv \int_1^{\infty}e^{-zt}t^{-m} dt$ is the exponential integral function. This non-Markovian  master equation [Eq.~(\ref{MasterEquation})] is an example of the general formalism developed in Ref.~\cite{aquino2017chemical}, see also~\cite{yin2021optimal}, and in the Appendix we include its full derivation. Note that, the first  term on the right-hand side of Eq.~\eqref{MasterEquation} is the contribution from the exponential reaction, while the second term emanates from the non-exponential birth with memory kernel $M(n, t)$. 

In the following we investigate master equation~(\ref{MasterEquation}) in two regimes: $\alpha>1$ and $\alpha<1$, and for various functional forms of $\tilde{\kappa}(n)$ that give rise to metastability.

\section{The case of $\alpha > 1$} \label{sec:largealpha}
We start with the case of $\alpha>1$, where we assume a functional form of $\tilde{\kappa}(n)$ such that the dynamics relax to a long-lived (quasi)stationary  state at long times. 
We start by simplifying the master equation \eqref{MasterEquation}. First, we Laplace-transform Eq.~\eqref{MasterEquation}, multiply by the Laplace variable $s$, and use the final value theorem of the Laplace transform, $\lim_{s\to 0} s \tilde{P}_n(s) = P_n$~\cite{yin2021optimal}, resulting in
\begin{equation} \label{SteadyMasterEquation}
    0 = (n+1) P_{n+1} - n P_n + \mathcal{M}(n-1) P_{n-1} - \mathcal{M}(n) P_{n}. 
\end{equation}
Here, $P_n$ is the (quasi)stationary PDF, and $\mathcal{M}(n)$ is obtained by taking the limit of $s \to 0$, corresponding to  $t\gg 1$, in $\tilde{M}(n, s)$ of Eq.~\eqref{memorykernelexplicit}. This procedure yields
\begin{eqnarray} \label{memorykernel_ss}
    \mathcal{M}(n) \!\!\!\!\!&& = \lim_{s \to 0} \tilde{M}(n, s)  \\ &&=  \tilde{\kappa}(n) \begin{cases}
         \frac{  E_{\alpha +1}\left(\frac{ \alpha n}{\tilde{\kappa}(n) }\right)}{E_{\alpha}\left(\frac{\alpha n}{\tilde{\kappa}(n)
   }\right)} + \OO[s/\tilde{\kappa}(n))]&  n > 0, \\  
   \frac{\alpha - 1}{\alpha} + \OO\{s/\tilde{\kappa}(n), [s/\tilde{\kappa}(n)]^{\alpha - 1}\} &  n = 0 . \nonumber
   \end{cases}
\end{eqnarray}
While the asymptotic result for $n\!>\!0$ holds for any $\alpha>0$, for $n\!=\!0$, it holds only for $\alpha>1$, and differs for different values of $\alpha$ (see also Sec.~\ref{sec:smallalpha}). Here, the notation $\OO\{s/\tilde{\kappa}(n), [s/\tilde{\kappa}(n)]^{\alpha - 1}\}$ is to be understand as $\OO\{s/\tilde{\kappa}(n)\}$ for $\alpha > 2$ and $\OO\{[s/\tilde{\kappa}(n)]^{\alpha - 1}\}$ for $1< \alpha < 2$. 
Yet, the state $n=0$ in Eq.~\eqref{memorykernel_ss} is negligible for $\tilde{\kappa}(n)\gg 1$ as the probability $P_0$ to be in the zeroth state is negligibly small, and thus it does not affect the overall dynamics. 

Defining $x = n/K$ and $\tilde{\kappa}(n) = K \kappa(x)$, where $K$ is the population's carrying capacity  and $\kappa(x)$ is a normalized production rate, a stationary mean-field equation for the average normalized population size, $\bar{x} \equiv \bar{n}/K$, reads
\begin{equation} \label{mean_field_equation}
    \bar{x} = m(\bar{x}), 
\end{equation}
where we have further defined
\begin{equation} \label{memorykernel_wkb}
    m(x) = K^{-1} \mathcal{M}(x/K) =  \kappa(x) \frac{  E_{\alpha +1}\left(\frac{x\alpha}{\kappa(x) }\right)}{E_{\alpha}\left(\frac{x\alpha}{\kappa(x)}\right)}. 
\end{equation}
For any specific choice of $\kappa(x)$, this deterministic mean-field equation can be solved analytically in simple cases (see examples below), or numerically, for the general case. 

To study escape from a metastable state we assume that the system at hand has (at least) two fixed points, one stable, and one unstable~\cite{assaf2010extinction}. Namely, Eqs.~(\ref{mean_field_equation}) and (\ref{memorykernel_wkb}) have (at least) two solutions. Accounting for stochasticity, the stable fixed point becomes metastable, and  our aim is to compute the mean time to escape from it.

Let us assume the initial population size is in the basin of attraction of (one of) the stable fixed points, which we denote by $\bar{x}_j$.  We are interested to compute the quasi-stationary PDF around this state, using Eq.~(\ref{MasterEquation}), and the mean time to escape from it.   
Starting with Eq.~\eqref{SteadyMasterEquation}, we substitute the WKB ansatz $P_n \sim e^{-K S(n/K)} \equiv e^{-K S(x)}$ into Eq.~\eqref{SteadyMasterEquation}, where $S(x)$ is called the action function, and neglect terms of order $\OO(1)$. This results in a stationary Hamilton-Jacobi equation, $H = 0$, with Hamiltonian 
\begin{equation}
    H = m(x) \left(e^{p}-1\right)+ x\left(e^{-p}-1\right), 
\end{equation}
where $p =  dS/dx$ is the conjugate momentum and $m(x)$ is given by Eq.~\eqref{memorykernel_wkb}. We note that  
Eq.~\eqref{mean_field_equation} for the average normalized population size, $\bar{x} \equiv \bar{n}/K$, is obtained by taking the limit $p\to 0$ in the Hamiltonian $H$. 
To obtain an expression for the action $S(x)$ we solve $H = 0$ for the momentum and integrate over $x$, which yields 
\begin{equation} \label{action}
    S(x) = \int_{\bar{x}_j}^x p(x') dx'  = -\int_{\bar{x}_j}^x \ln \left[ \frac{m(x')}{x'} \right] dx',
\end{equation}
where this integral can be computed for a given $\kappa(x)$. Having computed $S(x)$, we have the complete PDF of the long-lived metastable state around  $\bar{x}_j$, $P(x)\sim e^{-K S(x)}$. In particular, the variance of the PDF, $\sigma^2$, can also be computed. Using Eq.~\eqref{action}, we find:
\begin{equation} \label{var_general}
    \sigma^2 =\frac{K}{|S''(\bar{x}_j)|}= \frac{K \bar{x}_j m(\bar{x}_j)}{m(\bar{x}_j) - \bar{x}_j m'(\bar{x}_j)},
\end{equation}
where $\bar{x}_j$ is a stable  solution of Eq.~\eqref{mean_field_equation}. 

Even without specifying $\kappa(x)$, these results can be drastically simplified in the limit of  $\alpha \gg 1$, i.e., very weak memory. Here, we approximate the exponential integral function by Taylor-expanding around the lower limit of the integral: $E_m(z) \equiv \int_1^{\infty}e^{-zt}t^{-m} dt \simeq \int_1^{\infty}e^{-z -(z+ m)(t-1)} dt = e^{-z}/(m+z)$. Substituting this approximation into both exponential functions in Eq.~\eqref{memorykernel_wkb} and approximating for $\alpha \gg 1$ yields 
\begin{equation} \label{m_x_approx}
    m(x) \simeq \kappa (x)-\frac{\kappa (x)^2}{\alpha  (\kappa (x)+x)}. 
\end{equation}
The leading order is to be expected for exponential waiting times (memoryless process), while the second term is the leading correction in $1/\alpha$. 
In case of multiple fixed points in the language of the mean-field dynamics, fixed point $j$ is determined by solving
\begin{equation} \label{mean_field}
    \bar{x}_j = \kappa (\bar{x}_j )-\frac{\kappa (\bar{x}_j )^2}{\alpha  (\kappa (\bar{x}_j )+\bar{x}_j )}. 
\end{equation}
This equation can be solved perturbatively. In the leading order we denote the solution as $x_{j,0}$ such that $x_{j,0} = \kappa(x_{j,0})$. Next, we substitute the ansatz $\bar{x}_j = x_{j,0} + \delta x_{j}/\alpha$ into  Eq.~\eqref{mean_field} and solve for $\delta x_{j}$. This yields 
\begin{equation} \label{mean_field_sol_general}
    \bar{x}_j = x_{j,0} + \frac{x_{j,0}}{2\alpha   \left(\kappa '\left(x_{j,0}\right)-1\right)}, 
\end{equation}
which is consistent with the above ansatz.

While the expression for the action in Eq.~\eqref{action} holds for any $\alpha > 1$, it can be greatly simplified 
for $\alpha \gg 1$, using Eq.~(\ref{mean_field_sol_general}). 
Approximating the exponential integral function as above,  Eq.~(\ref{action}) becomes 
\begin{equation} \label{action_general}
    S(x) = -\int_{\bar{x}_j}^x \log \left[\frac{\kappa (x)}{x}\right] \, dx + \frac{1}{\alpha} \int_{x_{j,0}}^x \frac{\kappa (x)}{  (\kappa (x)+x)} \, dx,
\end{equation}
where $\bar{x}_j$ is a stable  solution of Eq.~\eqref{mean_field}. Here, in the second integral we approximated $\bar{x}_j$ by $x_{j,0}$; accounting for its ${\cal O}(\alpha^{-1})$ correction yields only $\OO[\alpha^{-2}]$ terms. To compute the mean time to escape from $\bar{x}_j$, we denote by $\bar{x}_i$ the adjacent unstable fixed point, and put $x=\bar{x}_i$ in Eq.~\eqref{action_general}, yielding the action barrier to escape~\cite{dykman1994large,assaf2010extinction,assaf2017wkb}:
\begin{equation} \label{action_general2}
    S\equiv S(\bar{x}_i) = -\int_{\bar{x}_j}^{\bar{x}_i} \!\!\log \left[\frac{\kappa (x)}{x}\right] \, dx + \frac{1}{\alpha} \int_{x_{j,0}}^{x_{i,0}}\!\!\frac{\kappa (x)}{  (\kappa (x)+x)} dx.
\end{equation}
Here, we have again substituted the zeroth order fixed points in the second integral to avoid corrections of 
$\OO[\alpha^{-2}]$. Equation~(\ref{action_general2}) provides a general solution for the (logarithm of the) mean time to escape from a metastable state, given a power-law WT with $\alpha\gg 1$ of the birth process. In this limit, the variance~(\ref{var_general}) around $\bar{x}_j$ becomes 
\begin{equation}
    \sigma^2 = \frac{\bar{x}_j \kappa(\bar{x}_j)}{\kappa(\bar{x}_j) - \bar{x}_j\kappa'(\bar{x}_j)} \left[ 1 + \frac{1}{\alpha} \frac{\bar{x}_j\kappa(\bar{x}_j)}{(\bar{x}_j + \kappa(\bar{x}_j))^2} \right],
\end{equation}
indicating that the variance increases as $\alpha$ is decreased, namely, as the  memory strength is increased.

Below we provide explicit results for the mean time to escape by considering several examples of $\kappa(x)$. 

\begin{figure}[t]
    \includegraphics[width=0.35\textwidth,clip=]{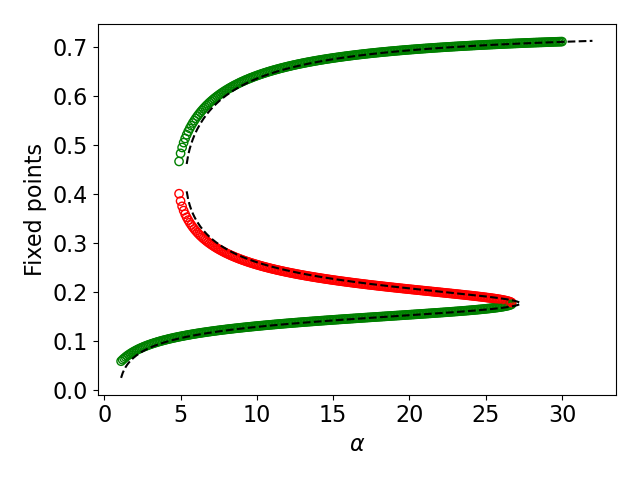}
    \vspace{-5mm}
        \caption{The stable and unstable fixed points in the case of phenotypic switching versus $\alpha$, with $\kappa(x)$ given by Eq.~\eqref{kappa_general}. We compare the numerical solution of $\bar{x} = m(\bar{x})$, see Eq.~\eqref{mean_field_equation} (empty circles), with the analytical approximation in the limit of $\alpha\gg 1$, given by Eq.~\eqref{mean_field_sol_general} (dashed lines). Stable and unstable fixed points are denoted in green and red, respectively. The parameters in Eq.~(\ref{kappa_general}) are $h=2$, $x_0 = 0.53$ and $f = 0.08$.}
    \label{fig:fig2}
\end{figure}

\subsection{Phenotypic switching} \label{sec:examples}
In the context of phenotypic switching, a relevant choice of $\kappa(x)=(1/K)\tilde{\kappa}(n)$ is a hill-type function, shown to be relevant in several biological systems, see e.g.,~\cite{roberts2011noise}:
\begin{equation} \label{kappa_general}
    \kappa(x) = f + \frac{x^h}{x^h+(x_0)^h}.
\end{equation}
Here, $x=n/K$ is the protein concentration,  $K\gg 1$ is the carrying capacity, while $h$, $f$, and $x_0$ are model parameters. The first term on the right corresponds to a constant flux and the second term gives rise to a positive feedback loop, for sufficiently large $h$. For exponentially-distributed WTs ($\alpha \to \infty$) 
genetic switching with hill-type production rate has been extensively studied, see e.g., Refs.~\cite{assaf2011determining, ge2015stochastic}. Here, the number of fixed points in the underlying deterministic dynamics depends on the parameters of Eq.~\eqref{kappa_general}. We focus on the case of two stable fixed points separated by an unstable fixed point, where stochastic fluctuations can drive the population to switch from one metastable state to the other.

We first study how memory affects the dynamical mean-field landscape by substituting Eq.~\eqref{kappa_general} into Eq.~\eqref{mean_field_equation}. Interestingly, we find that  molecular memory can lead to multiple bifurcations, which are demonstrated in Fig.~\ref{fig:fig2}. Here, we compare the approximate solution, Eq.~\eqref{mean_field_sol_general} with $\kappa(x)$ given by \eqref{kappa_general}, to the numerical solution of Eq.~\eqref{mean_field_equation}. One can see that for $\alpha \to \infty$ (no memory) there exists a single (high) stable fixed point for the chosen set of parameters. Similarly, for sufficiently low values of $\alpha$, there is a single (low) stable fixed point. However, for a wide intermediate range of $\alpha$ values, there are two stable fixed points, separated by an unstable point.

\begin{figure}[t]
    \includegraphics[width=0.48\textwidth,clip=]{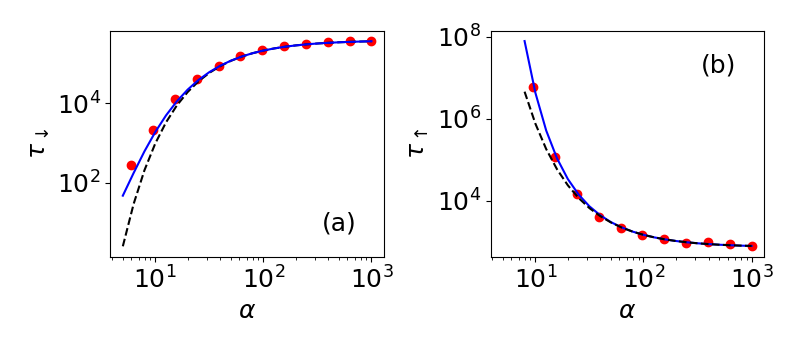}
            \vspace{-9mm}\caption{Mean time to switch between the metastable phenotypic states in a model of genetic switching, with $\kappa(x)$ given by Eq.~\eqref{kappa_general}. Left and right panels show the mean time to switch from the high to low and low to high states, respectively. Simulations (points) are compared with the numerical solution of \eqref{action} (solid line), and with the analytical approximation for $\alpha \gg 1$ [Eq.~\eqref{action_general2}]. Parameters are $h=2$, $x_0 = 0.49$ and $f = 0.06$. In (a) $N = 200$ whereas in (b) $N = 1500$.}
    \label{fig:fig3}
\end{figure}

In the latter regime, we further study the mean time to switch between the different metastable states, by substituting Eq.~\eqref{kappa_general} into Eq.~\eqref{action} for any $\alpha > 1$ and into \eqref{action_general2} for $\alpha \gg 1$. In both cases, the resulting integral can be solved numerically. In Fig.~\ref{fig:fig3}(a) we show an example of the mean time to switch from the high to low fixed points, $\tau_{\downarrow}$. The latter decreases with increasing molecular memory, i.e., decreasing $\alpha$. In contrast, the mean time to switch from the low to high states, $\tau_{\uparrow}$, increases with decreasing $\alpha$, as shown in Fig.~\ref{fig:fig3}(b). In both panels of Fig.~\ref{fig:fig3} we validate our results using numerical simulations.  Notably, we checked that the trends shown for both $\tau_{\downarrow}$ and $\tau_{\uparrow}$ are general for any choice of parameters that give rise to two stable fixed points, as was also shown in  Ref.~\cite{yin2021optimal} for a different WT distribution (Erlang function).



\subsection{Population establishment} \label{example:fixation} 
In the problem of population establishment, a population initially resides at a low fixed point with a few individuals, practically on the verge of extinction. Here, the quantity of interest is the mean time to escape from this low fixed point. To study this system under non-Markovian production, we take $\kappa(x)$ to be
\begin{equation} \label{kappa_fixation}
    \kappa(x) = f + \frac{x^2}{2}.
\end{equation}
Here we assume that creation or birth of new individuals  occurs due to influx of individuals and interactions between them, and is non-Markovian in nature~\footnote{Note that, Eq.~(\ref{kappa_fixation}) coincides with the small-$x$ expansion of Eq.~(\ref{kappa_general}) up to second order.}. 

\begin{figure}[t]
    \includegraphics[width=0.48\textwidth,clip=]{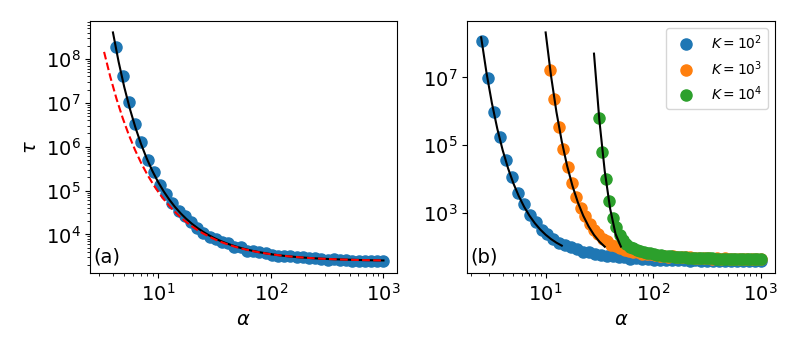}
            \vspace{-7mm}\caption{Mean time to population establishment  versus $\alpha$ for $f<0.5$ (a) and $f>0.5$ (b), for $\kappa(x)$ given by~\eqref{kappa_fixation}. Simulations (points) are compared to the numerical solution of \eqref{action} (solid line). In (a) we also plot the analytical approximation for $\alpha \gg 1$ [Eq.~\eqref{action_general2}]. The parameters are (a) $f = 0.43$ and $K = 100$ and (b) $f = 0.51$ and varying $K$ values (legend).}
    \label{fig:runaway}
\end{figure}

We start with the case of $0< f <0.5$, such that for $\alpha \to \infty$ there exists a stable fixed point of the underlying deterministic dynamics. For $\alpha\gg 1$, the stable and unstable fixed points are respectively given by 
\begin{equation} \label{fixation_fp}
 \bar{x}_1= (1 -\delta) \left(\!1 - \frac{1}{2  \delta \alpha }\!\right),\;\;\bar{x}_2= (1+ \delta)\left(\!1 + \frac{1}{2 \delta \alpha }\!\right),
\end{equation}
with $\delta = \sqrt{1-2f}$. Note that the distance between the fixed points increases as we decrease $\alpha$, in agreement with Fig.~\ref{fig:fig2} for the hill-type $\kappa(x)$.
The mean time for establishment is then estimated to be the time it takes the system to reach the unstable fixed point, starting from the vicinity of the stable fixed point. While the problem can be solved numerically for any $\alpha>1$, we provide analytical results for $\alpha \gg 1$. To do so, we substitute Eq.~\eqref{kappa_fixation} into Eq.~\eqref{action_general2} and integrate between the stable and unstable fixed points in Eqs.~\eqref{fixation_fp}. This results in 
\begin{equation}
    \tau = e^{K S}\;, \;\; S = S_0 + \alpha^{-1} S_1
\end{equation}
with 
\begin{eqnarray}
    S_0 &=& - 2 \sqrt{1-\delta ^2} \sin ^{-1}(\delta ) + 2 \delta\nonumber\\
    S_1 &=&  2 \delta- \delta^{-1}\log \left(1-\delta ^2\right) - 2  \tanh ^{-1}(\delta ),
\end{eqnarray}
where $S_0$ was computed previously, see e.g., Refs.~\cite{meerson2008noise,israeli2020population}. Here, the mean time for establishment grows as  $\alpha$ decreases, i.e., memory delays establishment. This is shown in Fig.~\ref{fig:runaway}(a), and agrees with Fig.~\ref{fig:fig3} above. 

Notably, for $f > 0.5$, in the limit of exponential rates (i.e., for $\alpha \to \infty$) starting from any initial condition leads to population runaway. That is, the dynamics are unstable and the population grows indefinitely. Yet, incorporating memory, $\alpha$ can act as a stabilizing mechanism, where only below some critical value of $\alpha$ the deterministic dynamics display a stable fixed point. This is shown in Fig.~\ref{fig:runaway}(b), where for a given choice of parameters (see caption), the fixed point appears at some $\alpha_c$, below which a low stable fixed point emerges, whose stability increases as the carrying capacity is increased. To estimate $\alpha_{c}$ below which the population is stable, we assume a-priori $\alpha_c \gg 1$, and use the Eq.~\eqref{m_x_approx} directly in the mean field equation \eqref{mean_field_equation}  for $f >0.5$. For $K\to \infty$ this yields
\begin{equation}
    \alpha_c = 2f/(2f-1). 
\end{equation}
This result is consistent with Fig.~\ref{fig:runaway}(b). Here, for $f \!=\! 0.51$ we find $\alpha_c \!=\!  51$, consistent with the largest-$K$ results. 

\subsection{Extinction of a bi-stable population} \label{example:extinction}
The third example we consider is an ecological example: extinction of an established population, under the Allee effect~\cite{stephens1999allee,mendez2019demographic}. Here $\kappa(x)$ satisfies~\cite{vilk2020extinction} 
\begin{equation} \label{kappa_extinction}
    \kappa(x) = \frac{x^2}{x^2 + x_0^2}  , 
\end{equation}
where $0<x_0<0.5$ is a model parameter. Note that this choice is equivalent to Eq.~\eqref{kappa_general} with $f =0$.
Here, in the limit of $\alpha \to \infty$, the dynamics give rise to three fixed points. These can be found analytically up to sub-leading order in $\alpha \gg 1$, by substituting Eq.~\eqref{kappa_extinction} into Eq.~\eqref{mean_field_sol_general}: 
\begin{eqnarray}
    &&\bar{x}_0 = 0, \nonumber
    \\
    &&\bar{x}_1 = (1/2)\left(1-\eta\right) + [1/(4\alpha)] \left(\eta^{-1}-1\right), \nonumber
    \\ 
    &&\bar{x}_2 = (1/2) \left(1+ \eta\right)-[1/(4\alpha)]\left(\eta^{-1}+1\right), 
\end{eqnarray}
with $\eta \equiv \sqrt{1 - 4x_0^2}$. 
Here, $\bar{x}_0$ is an absorbing stable fixed point, $\bar{x}_1$ is an unstable fixed point and $\bar{x}_2$ is a stable fixed point. Note that, the distance between the fixed points decreases as $\alpha$ is decreased, and a bifurcation occurs at some $\alpha_c$ when the fixed points coincide. Using the large-$\alpha$ approximation and equating $\bar{x}_1 = \bar{x}_2$ we find 
\begin{equation}
    \alpha_c \simeq 1/(2\eta ^2).
\end{equation}
Note that, this result is valid as long as $\alpha_c\gg 1$, i.e., $x_0$ must be close to $0.5$; as $x_0\to 0.5$, $\alpha_c$ tends to $\infty$. Yet,  for $\eta={\cal O}(1)$ the approximation breaks down, as $\alpha_c={\cal O}(1)$.

\begin{figure}[t]
    \includegraphics[width=0.35\textwidth,clip=]{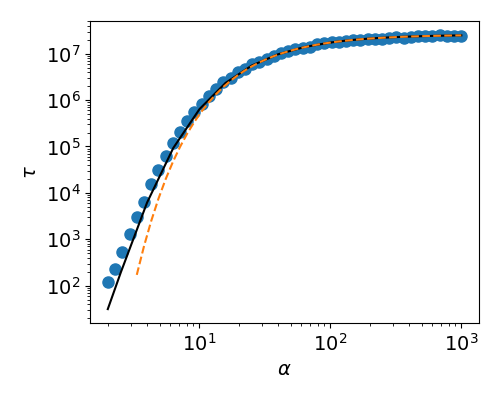}
            \vspace{-5mm}\caption{Mean time to extinction for $\kappa(x)$ given by Eq.~\eqref{kappa_extinction} with  $x_0 = 0.35$ and $K = 100$. Simulations (points) are compared to the numerical solution of \eqref{action} (solid line). We also plot the analytical approximation for $\alpha \gg 1$ [Eq.~\eqref{action_extinction}].}
    \label{fig:fig4}
\end{figure}

Apart from studying the mean-field dynamics, we also compute the mean time to extinction of the established population. To first order in $\alpha \gg 1$ we substitute Eq.~\eqref{kappa_extinction} into Eq.~\eqref{action_general2} and integrate between the stable fixed point $\bar{x}_2$ and the unstable fixed point $\bar{x}_1$~\cite{vilk2020extinction}. This yields: 
\begin{equation} \label{action_extinction}
    \tau = e^{K S}\;, \;\; S = S_0 + \alpha^{-1} S_1
\end{equation}
with 
\begin{eqnarray}
    S_0 &=& \eta- 2 x_0  \left[\cot ^{-1}\left(\frac{2 x_0 }{\eta-1}\right)+\cot ^{-1}\left(\frac{2 x_0 }{\eta+1}\right)\right]   \nonumber\\
    S_1 &=& - \tanh ^{-1}(\eta ) - \eta^{-1}\ln (2 x_0),
\end{eqnarray}
where $S_0$ has been computed in~\cite{meerson2008noise,assaf2010extinction,israeli2020population}. Note that the mean time to extinction significantly decreases as $\alpha$ is decreased, in agreement with the results in Fig.~\ref{fig:fig2}(a), indicating that memory destabilizes the high fixed point. In Fig.~\ref{fig:fig4} we compare the mean time to extinction given by a numerical solution of \eqref{action}, with the analytical approximation for $\alpha \gg 1$, Eq.~\eqref{action_extinction}, and numerical simulations, and find excellent agreement between the solutions.

\section{The case of $\alpha < 1$} \label{sec:smallalpha}
For $\alpha < 1$ the dynamics displays a markedly different behavior, as here the WT distribution between birth events given by $\psi_1$ [see Eq.~(\ref{psi1})] has a diverging mean. Recently, we have solved the case of $\tilde{\kappa}(n) = K$, i.e., a rate-independent production of the population size, which already gave rise to nontrivial results~\cite{vilk2023non}. Here we generalize the theory for a rate-dependent $\tilde{\kappa}(n)$, focusing on the effects of the initial conditions on the long-term dynamics. Considering $\alpha < 1$, the main difference from the description in Sec.~\ref{sec:largealpha} is the asymptotic form of the memory kernel \eqref{memorykernelexplicit} for $n = 0$. As mentioned above, for states with $n>0$ the memory kernel can be approximated by \eqref{memorykernel_ss} for any $\alpha>0$. However, for state $n = 0$ the memory kernel in the Laplace domain reads
\begin{eqnarray} \label{memorykernel_nzero_series}
   \hspace{-5mm} \tilde{M}(0, s) &=& \tilde{\kappa}(0)\,  E_{\alpha +1}\left(\frac{\alpha}{\tilde{\kappa}(0)}s\right)\bigg/E_{\alpha}\left(\frac{\alpha}{\tilde{\kappa}(0)
   }s\right)\nonumber\\
   &=&\frac{  \tilde{\kappa}(0)\left(\frac{\alpha  s}{\tilde{\kappa}(0) }\right)^{1-\alpha }}{\alpha  \Gamma (1-\alpha )}\left\{1 \!+\! \OO\left[\left(\frac{\alpha s}{\tilde{\kappa}(0)}\right)^{1-\alpha}\!\right]\!\right\}\!\!,
\end{eqnarray}
where in the second line we have taken the limit of $s\to 0$ corresponding to $t\to \infty$.
In contrast to Eq.~\eqref{memorykernel_ss}, here the leading term depends on the Laplace variable $s$. This is because at $n=0$ the only reaction that can occur is production, and since production is power-law distributed this gives rise to increasingly long times of inactivity. In contrast, for any $n>0$ the system will not be inactive since degradation can always occur. 

Taking the inverse Laplace transform of~\eqref{memorykernel_ss} and~\eqref{memorykernel_nzero_series}, after some algebra we obtain the memory kernel at $n\!=\!0$ 
\begin{equation} \label{memorykernel_nzero_series_time}
    M(0, t) \simeq \tilde{\kappa}(0)\begin{cases}
          \frac{(\alpha -1) }{\alpha } & \alpha > 1 ,\\ 
         \frac{(\alpha -1)\tilde{\kappa}(0)^\alpha \sin (\pi\alpha)}{\alpha^\alpha \pi t^{2-\alpha}} & \alpha < 1 .    
    \end{cases}
\end{equation}
For $\alpha < 1$ the dynamics display long-range correlation even at infinitely long times. This is a clear signature of nonergodicity~\cite{vilk2023non}, which entails significant variability across different realizations~\cite{metzler2000random}. To show this we derived an equation for the mean population size by multiplying Eq.~\eqref{MasterEquation} by $n$ and summing over all $n$. This yields 
\begin{equation} \label{nbar_protien_only}
    \pd{\overline{x}}{t} = - \overline{x}(t) + (1/K)\int_0^t \overline{M(n, t-t')}dt', 
\end{equation}
where $\overline{M(n, t - t')} \equiv \sum_{n=0}^{\infty} M(n, t-t')P_n(t')$.  This mean-field equation for the mean population size $\overline{x}$ cannot be solved in a straightforward manner. Nevertheless, for $\alpha > 1$ we showed above that 
$\overline{m(x)}\simeq m(\overline{x})$, see Eqs.~\eqref{mean_field_equation} and \eqref{memorykernel_wkb}, suggesting that for $\alpha > 1$ the dynamics are ergodic. In contrast, for $\alpha <1$ a single state retains memory much longer than any of the other states and the ergodic assumption does not hold~\cite{aquino2017chemical, metzler2014anomalous}. As shown in Ref.~\cite{vilk2023non}, at sufficiently long times, $t\gg \tau_0$ (to be defined below)
one obtains $\overline{n} \sim \overline{x}\sim t^{-(1-\alpha)}$, which is the long-time asymptotic of the dynamics for any $\alpha <1$.

\begin{figure}[t]
    \includegraphics[width=0.48\textwidth,clip=]{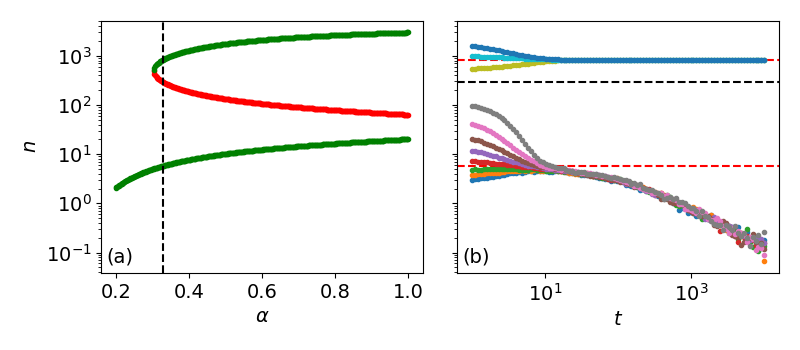}
            \vspace{-7mm}\caption{(a) Numerical solution of the average population size, $\bar{n}=K\bar{x}$, versus $\alpha$ (circles). Here $\kappa(x)$ is taken from Eq.~\eqref{kappa_general}, with $h=2$, $N=5000$, $f=0.005$ and $x_0=0.1$. Stable and unstable fixed points are denoted in green and red, respectively. The vertical dashed black line denotes $\alpha = 0.33$ used in panel (b). (b) Average population size (simulations) versus time for different initial conditions and similar parameters as in (a) with $\alpha = 0.33$. The dashed red lines represent the stable fixed points (high and low) and the dashed black line denote the unstable fixed point of the dynamics.}
    \label{fig:fig5}
\end{figure}

The decay of the population size is ultimately caused by the state $n = 0$, which is only visited at very long times when the dynamics are effectively stationary. In Ref.~\cite{vilk2023non}, for the linear case $\tilde{\kappa}(n) = K$, it was found that the typical time until asymptotic decay is proportional to the mean time to reach the absorbing state $n = 0$, denoted as $\tau_0$. At times $t< \tau_{0}$, the dynamics are effectively stationary, and  the average population size at times $t < \tau_0$, $\bar{n}_{ss}$, can be calculated using the stationary mean field equation. 

Here we perform a similar analysis and compute $\tau_{0}$, the mean first passage time to reach $n=0$. Using Eq.~\eqref{action} the latter satisfies, $\tau_0= \mathcal{N} e^{K S(0)}$, where the integral in~\eqref{action} is calculated numerically and $\mathcal{N}$ is some unknown pre-exponential factor. 
For general $\kappa(x)$ this description is still valid, but one needs to take into account the existence of multiple fixed points for the dynamics. For example, when $\alpha<1$ and $\kappa(x)$ is given by Eq.~\eqref{kappa_general}, for sufficiently low $\alpha$, only the low fixed point exists, see Fig.~\ref{fig:fig5}(a). Thus, the mean first passage time between the low fixed point and the ground state determines the typical time for decay. Yet, for larger values of $\alpha$, the typical time to decay will be determined by the maximum of $\tau_0$ and the mean time to switch from the high to low fixed points, $\tau_{\downarrow}$, see Sec.~\ref{sec:examples}. For instance, focusing on $\alpha = 0.33$ in Fig.~\ref{fig:fig5}(a) (vertical dashed black line), one can numerically find $\tau_0 \simeq 14$ (given the parameters of Fig.~\ref{fig:fig5}), whereas $\tau_{\downarrow} \simeq 1.3\cdot 10^{10}$. This behavior is demonstrated in Fig.~\ref{fig:fig5}(b). Here, starting from the vicinity of the low fixed point, the population decays relatively fast; yet, starting from the high fixed point, the population remains stationary and ergodic for a long time. This indicates that the long-term stationarity and ergodicity of the population strongly depend on the underlying dynamics, on the memory strength, as well as on the initial conditions.

\section{Conclusions} \label{conclusions}

We have studied several examples of ecological and biological single-population models in the presence of state-dependent birth and death rates, with delayed birth events due to a fat-tailed WT distribution. In order to study population escape, we focused on scenarios where the dynamics give rise to one or more metastable states. For distributions with a finite mean ($\alpha >1$) we derived results for the probability distribution of population sizes and found the mean escape time from a metastable state. In particular, we found that when memory is strong, corresponding to a small $\alpha$ value, switching from the high to low fixed points is expedited, while the reverse switch is delayed. This pattern is observed over a general class of models and parameters and has also been demonstrated for a different form of memory in the context of gene regulation, see Ref.~\cite{yin2021optimal}.
Furthermore, by studying examples of population extinction and establishment, we have revealed an effect of memory-induced stability, where for a sufficiently low $\alpha$ (sufficiently strong memory) a long-lived metastable state emerges.

When the WT distribution's mean diverges ($\alpha < 1$) we showed that the mean population size decays after a typical time, which scales as the mean first passage time to the ground state. Here, the dynamics are ergodic at short times but become nonergodic once the ground state is sampled~\cite{vilk2023non}. Notably, when the dynamics include multiple fixed points, there is a strong dependence on the initial condition, as the mean first passage time to the ground state can be markedly different when starting from the vicinity of the low or high stable fixed points. 

Finally, while we have focused here on single-population models relevant in ecology and cell-biology, an interesting future direction is to incorporate non-Markovian dyanmics and apply these techniques  in epidemiological models as well as multi-species competition models. Here, one expects to see similar dramatic effects on the system's dynamics under long-term memory.

\renewcommand{\thefigure}{A\arabic{figure}}    
\setcounter{figure}{0}

\renewcommand{\theequation}{A\arabic{equation}}    
\setcounter{equation}{0} 

\section*{Appendix}
Here we include a derivation based on Refs.~\cite{aquino2017chemical,yin2021optimal} of  master equation~\eqref{MasterEquation} and  memory kernel~\eqref{memorykernelexplicit}.

Using reactions~\eqref{reactions} we write the WT distribution for a single birth event to occur at time $t$ when the system has $n$ agents, given that no death occurs until time $t$:
\begin{equation} \label{phi2}
    \phi_1(n, t) = \psi_1(t) \int_t^\infty \psi_2(\tau) d\tau. 
\end{equation}
Similarly, the WT distribution for a single death event to occur given that no death occurs until time $t$ reads 
\begin{equation} \label{phi1}
    \phi_2(n, t) = \psi_2(t) \int_t^\infty \psi_1(\tau) d\tau. 
\end{equation}
We also define $R_k(n, t)$ as the joint probability density that the population size equals $n$ after $k$ reactions and $\Phi(n, t)$ as the probability density that no reactions occurred until time $t$. $R_k(n, t)$ obeys the recursion equation 
\begin{eqnarray} \label{R_recursion}
    R_{k+1}(n, t) = &&\int_0^t [R_k(n+1, t')\phi_2(n+1, t-t')  \\ &&+ R_k(n-1, t')\phi_1(n -1, t-t')] dt', \nonumber
\end{eqnarray}
where $R_0(n, t) = P_n(t)\delta(t)$, and $\delta(t)$ is the Dirac delta function. $\Phi(n,t)$ is given by 
\begin{equation} \label{Phi}
    \Phi(n,t) = 1 - \int_0^t[\phi_1(n, t') + \phi_2(n, t')] dt'. 
\end{equation}
Note that given Eqs.~\eqref{R_recursion} and \eqref{Phi} the PDF $P_n(t)$ obeys 
\begin{equation} \label{P_vs_R}
    P_n(t) = \int_0^t R(n, t') \Phi(n, t-t')dt', 
\end{equation}
where $R(n, t) = \sum_{k=0}^\infty R_k(n,t)$. 
Next, we Laplace transform  Eq.~\eqref{R_recursion} and sum the result over $k$, which yields
\begin{eqnarray} \label{R_laplace}
    \tilde{R}(n, s) = &&\tilde{R}_0(n, t) + \tilde{R}(n+1, s)\tilde{\phi_2}(n+1, s) \nonumber \\ &&
    + \tilde{R}(n-1, s)\tilde{\phi_1}(n-1, s), 
\end{eqnarray}
where we recall that $\tilde{f}(s)$ is defined to be the Laplace transform of $f(t)$, and $s$ is the Laplace variable. Taking also the Laplace transform of Eq.~\eqref{P_vs_R}, we obtain 
$\tilde{P}_n(s) = \tilde{R}(n, s)[1 - \tilde{\phi}_1 - \tilde{\phi}_2]/s$. Substituting this result into Eq.~\eqref{R_laplace} we find 
\begin{eqnarray} \label{P_recursion}
    s\tilde{P}_n(s) = && P_n(0) + \tilde{R}(n+1, s)\tilde{\phi}_2(n+1, s) \nonumber \\
    && + \tilde{R}(n-1, s)\tilde{\phi}_1(n-1, s) \nonumber \\
    && - \tilde{R}(n, s) [\tilde{\phi}_1(n, s)+\tilde{\phi}_2(n, s)]
\end{eqnarray}
Let us now define the memory kernels 
\begin{eqnarray} \label{M_12_def}
    \tilde{M}_i(n, s) = \frac{s\tilde{\phi}_i(n, s)}{1 - \tilde{\phi}_1(n,s) - \tilde{\phi}_2(n,s)}, 
\end{eqnarray}
for $i=1,2$. Using Eqs.~\eqref{M_12_def}  it is straightforward to show that $\tilde{M}_j (n,s)\tilde{P}_n(s) = \tilde{R}(n, s)\tilde{\phi}_j(n, s)$ for $j = 1, 2$. Substituting these equalities into Eq.~\eqref{P_recursion} we find 
\begin{eqnarray} \label{master_equation_laplace_app}
    s\tilde{P}_n(s) =  && P_n(0) + \tilde{M}_2(n+1, s) \tilde{P}_{n+1}(s) \\ 
    && + \tilde{M}_1(n-1, s) \tilde{P}_{n-1}(s) \nonumber\\ 
    && - [\tilde{M}_1(n, s) + \tilde{M}_2(n, s)] \tilde{P}_{n}(s) \nonumber
\end{eqnarray}
This is the generalized chemical master equation in the Laplace domain~\cite{aquino2017chemical, yin2021optimal}. Before taking the inverse Laplace transform, we compute the Laplace transformed memory kernels \eqref{M_12_def} by direct calculations. Substituting Eqs.~\eqref{psi1}, \eqref{phi2} and \eqref{phi1} into Eqs.~\eqref{M_12_def}  yields $\tilde{M}_2(n, s) = n$ and $M_2(n, t) = n \delta(t)$, while $\tilde{M}(n, s)\equiv\tilde{M}_1(n, s)$ is given by Eq.~\eqref{memorykernelexplicit} in the main text. Substituting $\tilde{M}_1$ and $\tilde{M}_2$ into Eq.~\eqref{master_equation_laplace_app} and taking the inverse Laplace transform yields Eq.~\eqref{MasterEquation} in the main text.

\bibliography{references}

\end{document}